# Generative AI Search Engines as Arbiters of Public Knowledge: An Audit of Bias and Authority

**Li, Alice**            The University of British Columbia, Canada | alice.li@ubc.ca
**Sinnamon, Luanne**     The University of British Columbia, Canada | luanne.sinnamon@ubc.ca

## ABSTRACT
This paper reports on an audit study of generative AI systems (ChatGPT, Bing Chat, and Perplexity) which investigates how these new search engines construct responses and establish authority for topics of public importance. We collected system responses using a set of 48 authentic queries for 4 topics over a 7-day period and analyzed the data using sentiment analysis, inductive coding and source classification. Results provide an overview of the nature of system responses across these systems and provide evidence of sentiment bias based on the queries and topics, and commercial and geographic bias in sources. The quality of sources used to support claims is uneven, relying heavily on News and Media, Business and Digital Media websites. Implications for system users emphasize the need to critically examine Generative AI system outputs when making decisions related to public interest and personal well-being.

## KEYWORDS
Generative AI; search engines; sentiment analysis; information quality; cognitive authority

## INTRODUCTION

The introduction of Generative AI (GenAI) Internet search tools constitutes the biggest shift in public access to information since Google and other second-generation search engines became available over 25 years ago. Systems such as Google Gemini, Microsoft Copilot, and Perplexity, mimic the familiar interaction paradigm of search engines: users submit a query (or prompt), and the system responds with information and links to Internet sources. Yet, these new systems function very differently by drawing upon large language models, such as ChatGPT, to generate narrative responses and engage in dialogue with information seekers, rather than directing them to relevant web pages. *Large language models* (LLMs) are systems trained on large datasets to predict the upcoming word or phrases based on preceding words and surrounding context (Bender et al., 2021). By designing systems able to synthesize information and make claims in direct response to people's queries, GenAI search engines are being set up as powerful gatekeepers and authorities in the public knowledge sphere, with the potential to surpass even Google's enormous influence (Haider & Sundin, 2019; Mager et al., 2023). Yet, these systems are developing so quickly that we know very little about how they work and the likely impacts on individuals and society.

There are many reasons to be wary of this situation. LLMs are extremely powerful, but prone to errors: generation of convincing misinformation (hallucination) (Bubeck et al., 2023; Ji et al., 2023; White, 2024), perpetuation and amplification of biases acquired through the model training process (Bender et al., 2021; White, 2024), dissemination of disinformation (Shah & Bender, 2022), and limited or incorrect attribution of sources (Kelly et al., 2023; Li et al., 2023; Liu et al., 2023; Urman & Makhortykh, 2023). Information quality issues can arise as information is decontextualized and repurposed. Decades of human-centered research on search engine use have identified consistent patterns that are likely to carry over to use of GenAI search engines: media and information literacy levels are low (Haider & Sundin, 2019); search engines are trusted and afforded cognitive authority (Huvila, 2013); cognitive biases influence information-based decisions (Azzopardi, 2021); and people are susceptible to inaccurate information (Rubin, 2019). Further, the ability of LLMs to interact in a human-like way and generate text seemingly imbued with meaning and understanding, while possessing no human traits or principles, creates additional vulnerabilities to users of these systems (Bender et al., 2021). In short, the potential is high for unethical and harmful outcomes of the uncritical and unregulated adoption of hybrid Generative AI search engines (GenAI SEs).

In this study we sought to understand how GenAI responses are designed in relation to topics and queries that represent issues of public concern and welfare. Given the responsiveness of LLM applications, such as ChatGPT, to style the presentation parameters, we reasoned that an analysis of the responses would reveal the goals of the system designers and inform our understanding of potential impacts on users. In consideration of the potential for system responses to influence users' perspectives, we were particularly interested in *sentiment bias*, reflected in skew in the emotional valence of responses (Huang et al., 2020; Tian et al., 2023) and in tactics used to present claims and establish authority. Findings contribute to a rapidly growing body of work on GenAI search systems, with the goal of increasing public awareness of their strengths and limitations, inform designers of areas for improvement, and encourage information researchers and professionals to continue to monitor developments and educate users.

We conducted an audit of public-facing and freely available ChatGPT and Generative Pre-trained Transformer (GPT) applications in search engines: Perplexity and Microsoft Bing Chat. While Bing Chat has since been renamed as Microsoft CoPilot, we retain the name used at the time of data collection (June 2023). In the sections to follow, we review prior work, outline the methods of data collection and analysis, and present findings in two phases: A quantitative comparison of three systems (ChatGPT version 3.5, Bing Chat and Perplexity), followed by an inductive analysis of the responses and sources provided by the latter two GenAI search engines. We then discuss the findings and conclude the paper. The present study expands on the initial findings that were previously published (Li & Sinnamon, 2023).

## PRIOR WORK

The rapid deployment of Generative AI powered search tools over the past 2 years has been accompanied by major media coverage and a burst of related research activity, much of it only available in pre-print. Terminology is still evolving, with researchers variously describing these systems as LLM-enabled or LLM-based search engines (Kelly et al., 2023; Wazzan et al., 2024); LLM-based chatbots (Urman & Makhortykh, 2023), AI agents (White, 2024) and Generative AI or Generative search engines (Capra & Arguello, 2023; Liu et al., 2023). We adopted the term Generative AI search engine (GenAI SE) to represent these hybrid systems, which combine LLMs with search results to generate responses and links to current Web content.

### Evaluation

A number of published papers have evaluated various inputs and features of GenAI SEs. Roegiest and Pinkosova (2024) examined the readability of training datasets of LLMs and GenAI SEs and found that Flesch-Kincaid Grade Levels varied across several datasets from grade 6 to 9, to grade 10-12, which has implications for suitability for different user populations. An exploratory study compared the number of sources provided by the GenAI SE Bing Chat and a standard search engine for the same search tasks and identified that Bing Chat returned significantly fewer links (4) of which only 2 were cited in the response text, as compared to DuckDuckGo (10 links per page) (Kelly et al., 2023). Another team of researchers audited the verifiability of citations in several GenAI SEs: Bing Chat, NeevaAI, Perplexity.ai, and YouChat (Liu et al., 2023). They found that about 52% of the generated sentences were supported by citations, and 75% of responses were fully supported by citations. Perplexity had more citations to support its sentences than Bing Chat, whereas Bing Chat was more closely paraphrasing from the cited websites compared to Perplexity. An audit study of political bias and censorship compared 3 GenAI systems: ChatGPT, Google Bard, and Bing Chat (Urman & Makhortykh, 2023). The study found a high level of refusal to answer in Google Bard for queries about Putin (90%) compared to other political figures (e.g., Biden), and reported that Bing Chat provided less false information compared to the other systems (Urman & Makhortykh, 2023). A limitation of existing audit studies is lack of adherence to best practices for algorithm audits (Roegiest & Pinkosova, 2024; Urman & Makhortykh, 2023). Studies that collect data at a single point in time may fail to capture broader patterns and dynamic changes in the algorithms and responses (Metaxa et al., 2021b). Given that LLM systems are not rule-based, but rather employ deep-learning algorithms, outputs may vary substantially over repeated interactions.

### Sentiment Bias

*Bias* in SEs is any type of systematic skew that causes harm (Friedman & Nissenbaum, 1996). Many types and sources of bias influence SEs, including those that arise from system design and user interactions (Baeza-Yates, 2018; Friedman & Nissenbaum, 1996; Kay et al., 2015). Prior research has identified search engine bias that discriminated based on gender and race and gender, for example (Kay et al., 2015; Metaxa et al., 2021a; Noble, 2018). In this study, we focus on the presentation bias in GenAI system outputs. The tendency for search system output to be influenced by the specific terms, tone and intent of the query has long been understood (Haider & Sundin, 2019; Tripodi, 2018). One manifestation of this is sentiment bias (Huang et al., 2020; Tian et al, 2023). For example, GPT-2 was found to generate text with more positive emotion for the profession "baker" compared to "accountant" (Huang et al., 2020). Sentiment bias can arise from social bias in the training data, interaction bias, or other sources (Baeza-Yates, 2018; Bender et al., 2021; White, 2024). Studies have found sentiment varied by topic in web search results (Demartini & Siersdorfer, 2010; Kazai et al., 2019; White, 2013). Health-related queries retrieved more confirmatory and positive highly ranked results (White, 2013); whereas controversial queries (e.g., genetic cloning) retrieved results with more negative-oriented emotion (Kazai et al., 2019). People demonstrate preferences (selection bias) for search results with more positive sentiment, such that sentiment bias can impact information behaviour (Kazai et al., 2019; White, 2013). These findings also raise questions about *confirmation bias*, which users prefer search results that echo or reaffirm their preexisting beliefs (White, 2013).

**Authority and quality in search results**
Search engines are trusted and familiar tools that have acquired the status of *cognitive authorities*: influential sources of expertise (Wilson, 1983; Haider & Sundin, 2019). They function as communicators of knowledge and play a role as cognitive authorities in persuading and influencing public opinion (Hovland et al.,1953). Huvila (2013) found that cognitive authority is afforded both to search systems and to the activity of searching, which may change as systems respond with answers rather than search results. Research has extensively explored how people determine source credibility (e.g., Arazy et al., 2017; Hilligoss & Rieh, 2008; Rieh, 2002; Sundin, 2011) with a consistent finding that people often rely upon heuristics (rules, cues, shortcuts) rather than in-depth analysis of features. Taraborelli (2008) attributes this to a reliance on predictive judgement of quality, based on reputation, rather than evaluative judgement. With respect to GenAI SEs, it is unclear if and how these systems will establish authority, although many users are likely to uncritically accept them as next-generation search engines. Early research suggests that GenAI SE outputs appear to be fluent and coherent (Bender et al., 2023; Zelch et al., 2024), and informative (Capra & Arguello, 2023; Liu et al., 2023; Zelch et al., 2024). Attribution, the capacity of a system to provide evidence and support for claims, for example, through citation of sources, is an important aspect of authority, and, given the nature of LLMs, is one of the challenges currently facing the research community (Li et al., 2023). This study will explore attributions and rhetorical features associated with authority.

**User perception of generative AI search engines**
A small number of user studies of these new systems have been published, including university students and personnel (Capra & Arguello, 2023; Wazzan et al., 2024) and crowdsourced workers (Liu et al., 2023; Malaviya et al., 2023; Zelch et al., 2024). Capra and Arguello (2023) conducted a small-scale study using a prototype LLM-based search system and reported that participants found the generated text fluent and informative; similar results were observed in the crowdsourced studies (Liu et al., 2023; Zelch et al., 2024). Another general observation across these studies is that participants were not aware of how information was generated (Capra & Arguello, 2023; Zelch et al., 2024), such as information about stocks (Capra & Arguello, 2023) or information that includes advertisement (Zelch et al., 2024). Zelch et al. (2024) called attention to *native advertisement* which they used to describe incidences of subtle product and brand placement in the generated text. Students had different levels of trust of the generated responses on informational content (Capra & Arguello, 2023), which may be attributed to the critical thinking skills gained in university. User studies are challenging to conduct, as these systems are evolving so rapidly, and practices associated with such systems are not yet established. While it is likely that established patterns of interaction with SEs will carry over, trust in AI is a subject of intense debate in society, which may influence uptake and use of these systems at least by some, as suggested by Capra and Arguello's (2023) early study.

**Current study**
As this field is new, there are many open questions. Prior research indicates a need to invest research effort in auditing the sentiment of GenAI systems using best practices and investigating the nature of the responses, including the ways in which claims are presented and supported. We have identified the following research questions:

RQ1: Are Generative AI responses influenced by the sentiment of the query and topic?
RQ2: How do Generative AI search engines establish authority in their responses to queries?

**RESEARCH METHODS**
We designed the study as an *algorithmic audit*: a method that systematically queries an algorithm, observing the responses to examine how the algorithm works (Sandvig et al., 2014). The method is derived from traditional auditing practices in the social sciences that examine human discrimination against individuals and groups, but with a focus on automated processes (Metaxa et al., 2021b). We conducted a *scraping audit*, in which repeated queries are used to observe system responses (Sandvig et al., 2014). Algorithmic audits are designed around two axes: the topic and the type of discrepancy (Metaxa et al., 2021b). To probe GenAI systems on bias and authority, we identified four topics representing current global issues: climate change, vaccination, alternative sources of energy, and trust in the media (Edelman, 2023; Government of Canada, 2018). Public knowledge and perspectives on these issues vary substantially among the general population and are significant in shaping policy and behaviour. The type of discrepancy examined in the first stage of the analysis is sentiment bias. We operationalized sentiment bias as a correlation between query sentiment polarity (positive, negative, neutral) and response sentiment polarity. The second stage of our analysis was limited to the responses of the two GenAI SEs (Bing Chat and Perplexity), which, unlike ChatGPT, are designed to function as alternatives to traditional SEs and include citations to sources. By analyzing the responses collected through the audit, we further explored their nature with respect to several additional variables (Table 2).

**Systems**

We selected three GenAI systems freely available for public use: GPT-3.5 version powered ChatGPT and Perplexity (AI team, 2023; OpenAI, 2023b) and GPT-4 powered Bing Chat (Mehdi, 2023). These systems are linked through the shared use of ChatGPT LLM. Early versions of Perplexity also relied on the Bing index, but all three systems have developed rapidly since they were released, and their precise implementations are proprietary and not known. Perplexity and Bing Chat both use retrieval augmented generation (RAG) to integrate LLM and search index components and include attributions to source references in system responses (Smith, 2024; White, 2024). At the time of data collection, Google's LLM-based system (formerly Bard; now Gemini) was not available in the country where this research was conducted, and therefore, was not included in the study.

**Topics and Queries**

We use the term "query" to collectively refer to prompts and queries. For each of the four topics, we gathered a set of 12 authentic queries from AnswerThePublic, a fee-based Web service that scrapes popular searches from Google autosuggestion data (NP Digital, 2023). From among the most frequent English-language queries originating in Canada in May 2023, we selected a set of queries with substantial sentiment variation within each topic. The number of queries used in this study aligns with other search engine algorithm audits (Hussein et al., 2020; Le et al., 2022). Table 1 includes examples of queries for each topic, including their sentiment polarity scores.

| Topic | Query | Sentiment Polarity |
|---|---|---|
| Climate Change | climate change benefit | 0.76 |
| | will climate change kill us | -0.57 |
| | will climate change ever stop | -0.06 |
| Vaccination | vaccination benefits | 0.58 |
| | why vaccine mandates are unethical | -0.61 |
| | vaccination in pregnancy | -0.08 |
| Alternative Energy | renewable energy is important | 0.84 |
| | why alternative energy is bad | -0.75 |
| | alternative energy advantages and disadvantages | 0.04 |
| Trust in Media | trust in media by country | 0.37 |
| | distrust in media | -0.72 |
| | trust in media all time low | -0.10 |

**Table 1. Example queries and sentiment polarity scores (-1 to +1).**

**Data Collection**

Best practices for algorithmic audits include conforming to legal and ethical guidelines regarding data scraping (Metaxa et. al., 2021b). While we found no guidelines for Bing Chat (Bing, 2023), Perplexity and ChatGPT disallow automatic data scraping (OpenAI, 2023a; Perplexity, 2023). Therefore, we collected the data through manual query submission and documentation of responses and sources. For seven days during one week in June 2023, 12 search queries for each topic (total 48 queries) were submitted daily to each of the three systems: Bing Chat, Perplexity, and ChatGPT. The study was conducted in Canada. The rationale for collecting responses over multiple days was to anticipate dynamic changes in the algorithms and responses, as the goal of this audit was not to target specific phenomena at a specific time, but broader trends (Metaxa et al., 2021b). Dummy accounts were used for ChatGPT that required sign-in, and data was collected via different public computers on a university campus to avoid effects of personalization. Default settings were used for each system to mimic non-expert information seeking behaviour. Although these systems invite conversational style interactions, we issued a single query and recorded the first response to it, without iterations. A fresh session was initiated after each interaction to reduce impacts of session memory.

**Data Preparation and Analysis**

There were two phases of this study: automated analysis, followed by response and source analyses. Table 2 presents the variables, measures, and sources. In the first phase of the analysis, Python (version 3.10) in the Spyder integrated development environment was used to pre-process the collected data ($N$=1,008) and to calculate response length and sentiment polarity scores of queries and responses for the three GenAI systems (ChatGPT, Bing Chat, Perplexity). Data cleaning was carried out using Regular expression (version 3.10.12; Python Software Foundation, 2023), Pandas (version 1.5.3; McKinney, 2010), and NumPy (version 1.25.0; Harris et al., 2020) packages. Sentiment analysis was computed using a fine-tuned transformer model (Yuan, 2023). In most cases, system responses to the same queries differed across the 7 days, but a small number of duplicate responses by Bing Chat were retained in the analysis, as they were collected on different days and by different accounts. Statistical analyses of sentiment polarity were conducted using IBM SPSS (version 29.0.1) to assess correlation and group differences by topic and search system.

The second phase focused on the responses of GenAI SEs: Bing Chat and Perplexity ($N$=672). Readability of responses was assessed using the Flesch Kincaid Grade Level metric (Kincaid et al., 1975), calculated using the textstat package (version 0.7.3, Bansal & Aggarwal, 2022) in Python. We then manually analyzed responses following a general inductive coding approach as outlined by Thomas (2006). The primary objective of the inductive analysis was to identify rhetorical tactics used to support claims in relation to the query and to establish authority. The two authors independently carried out open coding on a purposive sample of 30 responses that varied by sentiment polarity, topic, and system. We then met to compare outcomes and jointly devised a coding framework around 4 identified parameters: narrative voice, stance on query, and techniques to support and to mitigate claims. Codes and definitions are presented together with the results in Table 4. The unit of analysis was set at the response level. The first author coded all responses and a coder who was not involved in the project coded a 10% random sample of responses ($n$=67) (O'Connor & Joffe, 2020). Following an initial low inter-coder reliability score ($α$=0.59) (Krippendorf, 2018), the coders met, identified the main area of disagreement and revised the codebook for greater clarity. A subsequent coding round resulted in a reliability score of $α$=0.71, which is an acceptable strength of agreement (Krippendorff, 2019). Calculations of reliability scores were carried out in Python using the Fast Krippendorff package (version 0.6.1.; Castro, 2023). Results of the readability and rhetorical analyses are summarized using descriptive statistics and elaborated in a narrative reporting of results.

| Variable | Measure | Source of Measure |
|---|---|---|
| *Phase 1: ChatGPT 3.5, Bing Chat and Perplexity* | | |
| Query sentiment | Polarity score (-1 to +1) | Python 3.10, a transformer model |
| Response sentiment | Polarity score (-1 to +1) | Python 3.10, a transformer model |
| Length of response | Word count | Python 3.10 |
| *Phase 2: Bing Chat and Perplexity* | | |
| Readability of response | Flesch Kincaid Grade Level | Python 3.10, Textstat package |
| Rhetorical features | 4 feature types and values (details in Table 4) | Inductive coding, manual |
| Number of citations | Source count | MS Excel |
| Source domain type | 14 Source type categories | Manual classification |
| Source origin | Country or Region | Manual classification |

**Table 2. Study variables, measures, and sources of measures**

The sources cited in responses ($N$=3448) were captured and analyzed at the level of the web domain (e.g., www.cdc.gov), rather than the specific page level. We manually reviewed the 355 unique web domains to identify the geographic origin and to classify the source by type of media or organization. Using an inductive and iterative process we identified 14 categories of sources from the data, for example, News and Media, Commercial Business, and Government, and assigned each domain to a single category. The full category list is included in Table 5 in the results section. We then aggregated domains by system and by topic to conduct a quantitative descriptive analysis.

**RESULTS**

A descriptive summary of the 1008 responses by system and topic is presented in Table 3. The average length of results across all topics and GenAI systems is 234 words. Within each system, response length does not vary substantially across topics, but there is a notable difference across systems. ChatGPT responses are the longest ($M$=336), followed by Perplexity ($M$=243) and Bing Chat ($M$=123). The number of cited sources for Bing Chat and Perplexity responses is remarkably consistent across systems and topics at just over 5 sources per response.

| | ChatGPT* | | Bing Chat | | | Perplexity | | |
|---|---|---|---|---|---|---|---|---|
| **Topic** | Length *M(SD)* | Length min/max | Length *M(SD)* | Length min/max | Sources *M* | Length *M(SD)* | Length min/max | Sources *M* |
| **Climate change** | 347 (85) | 189/500 | 139(51) | 56/291 | 5.19 | 283(96) | 156/597 | 4.96 |
| **Vaccination** | 297(115) | 67/493 | 117(42) | 35/224 | 5.10 | 225(71) | 88/428 | 5.25 |
| **Alternative energy** | 346(90) | 114/528 | 117(35) | 49/227 | 5.02 | 214(50) | 90/426 | 5.27 |
| **Trust in media** | 353(82) | 91/635 | 121(34) | 62/260 | 5.19 | 250(58) | 149/505 | 5.06 |
| **Total – all topics** | 336(96) | 67/635 | 123(42) | 35/291 | 5.13 | 243(78) | 88/597 | 5.14 |

Table 3. Response length (words) and number of sources per response by system and topic.
*No sources were cited

## Sentiment Bias
*Query Sentiment Bias*
Preliminary analyses ($N$ =1,008) show the relationship between query polarity and GenAI system response polarity is linear with both variables normally distributed, as assessed by Shapiro-Wilk's test ($p$ >.05), and there are no outliers. Thus, a parametric Pearson's product-moment correlation is utilized to assess the relationship. The analysis identifies a moderate positive correlation between query polarity and response polarity, (1,006) = .46, $p$ < .001, with query polarity explaining 21% of the variation in response polarity. Similar levels of correlation are found for all three systems: ChatGPT ($r$ = .435, $p$ <.001), Bing Chat ($r$ = .454, $p$ <.001), and Perplexity ($r$ = .509, $p$ <.001).

*System and Topical Sentiment Bias*
Shapiro-Wilk's tests show non-normal distributions of response polarity for all three systems and topics ($p$ < .001), so we use non-parametric Kruskal-Wallis H tests to test for differences. Median response polarities are significantly different between groups for the three systems, $\chi^2(2)$ = 12.54, $p$ = .002, and the four topics, $\chi^2(3)$ = 299.36, $p$ < .001. Pairwise comparisons are conducted using Dunn's procedure, with a Bonferroni correction. Adjusted p-values are presented. Perplexity has significantly lower polarity of responses ($Mdn$ = -.19) than ChatGPT ($Mdn$ = -.01, $p$ = .05) and Bing Chat ($Mdn$ = -.10, $p$ = .05). Climate change ($Mdn$ = -.33) and trust in media ($Mdn$ = -.39) have significantly lower polarity than vaccination ($Mdn$ = .25, $p$ = .001) and alternative energy ($Mdn$ = .24, $p$ = .001).

## Response Analysis
We conducted a deeper analysis and comparison of the responses provided by the two GenAI SEs, Bing Chat and Perplexity. The mean Flesch Kincaid Grade Level is higher for Perplexity ($M$=12.82; $SD$=2.00) than Bing Chat ($M$=9.87; $SD$=2.11), indicating that Perplexity responses are designed for a higher educational reading level (post-secondary) compared to Bing Chat (secondary school). Beyond readability, we conducted a rhetorical analysis of responses, which is summarized in Table 4. For categories 1 and 2, responses were assigned a single code or no code if none applied. In categories 3 and 4, all instances of tactics were coded in each response.

Findings for narrative voice (1) indicate that Bing Chat tends to respond in the first person, for example, often adding, "I hope this helps." Perplexity often uses the third person and passive voice, with most responses referring to "search results" or to content from search results as the authority upon which the response is based.

In relation to the query (2), most responses in both systems support explicit or implicit claims in the query. For example, in response to, *will renewable energy stocks go up*, Bing Chat states, "Renewable energy stocks have been growing significantly for the last decade and are anticipated to continue this trend." Perplexity's response to *vaccination pros and cons* is supportive, and more controversial: "Experts who are against immunization disagree and actually find vaccines to be harmful to health because they cause adverse side effects." Cases of query opposition are limited (10-12%), and primarily occur in response to counter-factual queries, such as, *climate change is not caused by humans* and *why alternative energy is bad*. The two systems take different approaches when disagreeing. For example, some Bing Chat responses to *why vaccine mandates are unethical* note that it is controversial topic with different opinions, while others clearly state that mandates are ethical, citing health authorities. Perplexity responses to this query simply identified it as a controversial topic.

| Feature | Definition and example | Bing Chat (*n* = 336) | Perplexity (*n* = 336) |
|---|---|---|---|
| **1) Narrative voice of response (mutually exclusive)** | | | |
| First person singular | Makes claims as individual author - "I" | 81.0% | 3.9% |
| First person plural | Makes claims as group author - "we" | 5.4% | 10.1% |
| Third person | Attributes claims to other authorities - "search results", passive voice | 13.6% | 86.0% |
| **2) Stance in relation to Query (mutually exclusive)** | | | |
| Support | Agrees with/support viewpoint expressed in query | 81.3% | 81.0% |
| Neutral | Presents a balanced viewpoint in relation to the query | 6.8% | 8.9% |
| Oppose | Disagrees with/opposes viewpoint expressed in query | 11.9% | 10.1% |
| **3) Techniques used to support claims (non-mutually exclusive)** | | | |
| Generic authorities | References general authorities, e.g., experts, scientists | 56.2% | 56.0% |
| Named authorities | referenced specific named entities, e.g., NASA | 53.3% | 53% |
| Quantification | Numerical data and/or statistics are referenced | 51.2% | 45.2% |
| **4) Techniques used to mitigate claims (non-mutually exclusive)** | | | |
| Balance check | Offers a counterpoint or alternative perspective (e.g., "Some people believe... However, others believe...") | 25.9% | 39.0% |
| Conditional language | Notes that situations differ based on time, location, policy (e.g., "Different factors ... contexts can influence...") | 14.6% | 16.7% |
| Hedging | Uses language to add uncertainty (e.g., can, could, might, may, potential) and reduce specificity (e.g., some, many) | 74.1% | 83.6% |
| Notes limitations | Mentions shortcomings of its own knowledge (e.g., as an LLM, I cannot...) | 0.3% | 2.4% |

**Table 4. Summary of Rhetorical Features and Frequencies in Responses of Bing Chat and Perplexity**

To support their claims (3), more than half of the responses in Bing Chat and Perplexity refer to generic authorities, often "researchers" or "scientists" and/or named authorities. Perplexity frequently referenced vague, majority or minority viewpoints, such as "most people" or "some people". In response to the query *vaccination pros and cons*, the system offers both perspectives: the pro arguments tend to start with assertions such as "vaccines can save..." "vaccines help...", the con arguments start with "some people...". It is unclear if such subtle techniques of differentiation are picked up in the training data or designed into the response generation process. In some cases, named authorities are companies or products. For instance, Bing Chat recommended stocks (e.g., NextEra Energy) in response to the query *will renewable energy stocks go up*, and Perplexity frequently mentioned a specific company (Alternative Energy Solutions, Ltd.) when asked about *alternative energy solutions*.

Techniques to mitigate claims (4) are used extensively by both systems, with hedging being the most common. For instance, "Based on the search results, it is difficult to predict with certainty whether renewable energy stocks will go up. However, there are several factors that could influence the performance of these stocks." Notably, we found very few instances in which the systems admitted the limits or shortcomings in relation to these topics.

**Source Analysis**

Responses to the 672 (48 unique) queries issued to the two systems cited 3,448 sources in 355 unique web domains. Of these domains, 28% were cited only by Bing Chat; 46% were cited only by Perplexity, and 26% were cited by both. 25% were in the long tail: cited only once across the whole dataset, even though each query was submitted 7 times to each system. The average number of citations per response was equivalent for the two systems at 5.1 (5.125 Bing, 5.137 Perplexity). While sources ranged from 1 to 7 for the two systems, the majority had between 4 to 6. Sources originated from 24 countries, mostly in North America and Europe, but including Japan, Pakistan, Hong Kong, South Africa, and Australia. 95% of source domains originated in the United States (U.S.), United Kingdom (UK), Canadian, and international organizations (Figure 1). Overall, 65% of references were to sources in the U.S., even though queries were issued from Canada. Bing referenced more sources from the UK and Canada and fewer from the U.S. than Perplexity.

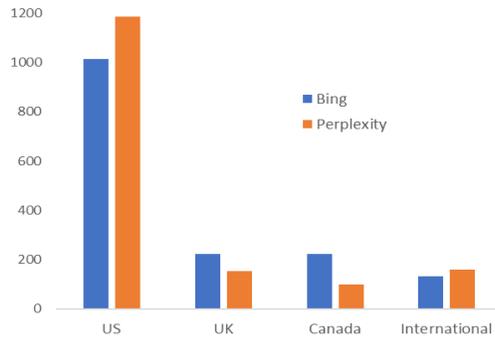

**Figure 1. Number of sources from the most frequently cited regions**
(*N*= 3190; 95% of total source domains)

Table 5 shows the distribution of source types by web domain, classified into 14 categories. Source types are listed in decreasing order of frequency and those greater than 10% are bolded. The most frequently cited category is News and Media (21.8%), which includes mainstream news and broadcasting organizations, followed by Business (17.9%), Government (16.5%), Non-Profit (13.3%) and Digital Media (11.7%). Digital media sites are news and information websites focused on specific domains and they range widely in quality. Nine of the 14 categories are cited infrequently, including several typically associated with high quality information: educational institutions (2.6%), academic publishers and journals (3.0%), policy institutes (1.7%) and professional and scholarly associations (3.1%).

|  | Bing Chat | | | | | Perplexity | | | | | |
|---|---|---|---|---|---|---|---|---|---|---|---|
| **Source Types** | **CC** | **Eng** | **Med** | **Vacc** | **Total** | **CC** | **Eng** | **Med** | **Vacc** | **Total** | **Total** |
| News and Media | **39.9** | **11.0** | **35.3** | **18.3** | **26.2** | **22.6** | 10.0 | **29.6** | 6.7 | **17.4** | **21.8** |
| Business | 0.5 | **40.0** | **15.4** | 2.3 | **14.3** | 4.6 | **43.6** | **25.7** | **10.7** | **21.6** | **17.9** |
| Government | **25.0** | 2.4 | 3.3 | **36.2** | **17.0** | **32.2** | 7.0 | 0.0 | **26.9** | **16.1** | **16.5** |
| Non-Profit | 8.3 | 2.6 | **23.0** | 6.7 | **10.1** | **17.5** | **11.1** | **18.1** | **19.5** | **16.5** | **13.3** |
| Digital Media Site | **11.5** | **21.1** | 7.1 | **20.6** | **15.1** | 6.3 | 5.5 | 3.6 | **18.7** | 8.3 | **11.7** |
| NGO | 6.9 | **14.1** | 0.0 | 6.4 | 6.8 | 4.8 | **11.1** | 0.0 | 4.0 | 5.0 | 5.9 |
| Association | 2.5 | 1.2 | 0.0 | 2.1 | 1.5 | 8.4 | 4.1 | 1.6 | 5.0 | 4.7 | 3.1 |
| Acad. Publisher | 0.0 | 1.7 | 9.5 | 0.9 | 3.0 | 0.5 | 0.2 | 6.1 | 5.2 | 3.0 | 3.0 |
| Educational Inst. | 2.1 | 4.3 | 0.0 | 3.9 | 2.6 | 0.7 | 4.1 | 3.4 | 2.0 | 2.6 | 2.6 |
| Policy Institute | 0.0 | 0.2 | 4.5 | 0.0 | 1.2 | 0.2 | 1.4 | 7.0 | 0.0 | 2.2 | 1.7 |
| Social Media Site | 1.6 | 0.7 | 1.9 | 0.7 | 1.2 | 1.9 | 0.2 | 4.3 | 0.0 | 1.6 | 1.4 |
| Personal Website | 0.9 | 0.0 | 0.0 | 1.6 | 0.6 | 0.0 | 0.0 | 0.0 | 0.0 | 0.0 | 0.3 |
| Search Engine | 0.0 | 0.7 | 0.0 | 0.0 | 0.2 | 0.0 | 0.0 | 0.0 | 0.0 | 0.0 | 0.1 |
| Other | 0.9 | 0.0 | 0.0 | 0.2 | 0.3 | 0.2 | 1.6 | 0.7 | 1.2 | 0.9 | 0.6 |
| Grand Total | 100% | 100% | 100% | 100% | 100% | 100% | 100% | 100% | 100% | 100% | 100% |

**Table 5. Distributions of Source Types (%) by System and Topic (CC=Climate Change, Eng=Renewable Energy, Med = Trust in Media, Vacc=Vaccination); >10% in bold**

Examining the sources by system reveals some differences. Bing Chat cites a greater proportion of sources from News and Media, and Digital Media domains than Perplexity, which cites relatively more Business and Non-Profit source domains. There are clear patterns in source citation across topics. In both systems, the Climate Crisis and Vaccination topics cite Government sources heavily; Business sources are most frequently cited for the Renewable Energy topic; and News and Media and Non-Profit sources are most frequently cited for Trust in Media.

With respect to information quality, several questions arise from this analysis, including the frequent citation of Digital Media Sites by Bing Chat, particularly for the Vaccination topic (20.6%) and the prevalence of Business sources, particularly for the Renewable Energy topic (>40%). In the first case, sources include popular consumer and medical health websites, including WebMD, Healthline, and VeryWell Health as well as some less authoritative sites, such as Grunge, rated as "questionable" by Media Bias/ Fact Check (2024). In the second case, for the renewable energy topic, Perplexity cites approximately 30 different company websites (e.g., Inspire Clean Energy) that offer a range of energy services, many of which include blog posts and articles about renewable energy, as well as finance, investment, and employment sites related to the energy sector. Across these 355 domains, we identified other questionable and poor-quality sources, including essay writing services (e.g., IvyPanda), a high school biology assignment, climate change denial articles, and advertising and content farms. While the most frequently referenced domains are relevant and authoritative, including government and non-governmental agencies (e.g., the US Center for Disease Control, Government of Canada, World Health Organization), research organizations (e.g., Gallup Consulting and Pew Internet Research, and major news agencies (e.g., British Broadcasting Corporation), this level of quality is not consistent.

## DISCUSSION

In response to RQ1, we found a significant impact of both query bias and topical bias in the GenAI system responses. A medium strength correlation was found between query and responses suggesting that responses echoed the emotional valence of the query. While cause and effect cannot be inferred, we varied the query polarity within each topic, so topic or domain effects are not likely to be factors. Results with respect to query stance, showing that most responses supported or were neutral in relation to the query, lend further support to the conclusion that responses are influenced by the wording and tone of the query, as has been observed in studies of traditional SEs (Tripodi, 2018). Query sentiment and stance bias may serve to limit the diversity of perspectives encountered by searchers and reinforce confirmation bias (Azzopardi, 2021). Topical bias was reflected in more positive sentiment for solution-focused topics (vaccination and alternative energy) over problem-focused topics (climate change and trust in media). This is not unexpected and likely reflects the sentiment in the model and the underlying web content, in keeping with results of prior research (Demartini & Siersdorfer, 2010; White, 2013).

In response to RQ2, we presented various indicators of authority and quality in GenAI SE responses. An interesting difference arises from the choice of first-person narration by Bing Chat and Perplexity's use of third person and passive voice. This reflects a different approach to establishing cognitive authority, which can be situated in different actors, elements or even activities (Hovland et al, 1953; Huvila, 2013). Bing Chat adopts the role of authoritative knower and "communicator," while Perplexity presents as an intermediary, passing on knowledge from search results and authoritative others. The second approach more faithfully represents the hybrid, RAG, model used by GenAI SEs, in which the LLM is prompted with search results and constructs a response from them. It remains to be seen if these different rhetorical approaches will shape users' mental models and impact perceptions of trust and credibility.

These responses are clearly designed to serve as authoritative, evidence-based sources of information for these topics. Both systems use a range of common techniques to support claims, including reference to specific and general authorities and statistics. At the same time, they make heavy use of rhetorical hedging and equivocation, which reduces the strength of arguments, and perhaps more importantly for these systems, limits the responsibility and reputational risk of the information providers. Both systems employ balance checks and surface minority voices ("some people") to present different perspectives, even for topics with clear factual grounding in science. Balance checks are associated with pseudo-science and the problematic media legacy of prioritizing apparent neutrality and balance over scientific consensus in issues such as climate change (Boykoff & Boykoff, 2004), so it is disturbing to see such techniques on display. As noted by Ghezzi and colleagues (2020), it is unethical for SEs to provide anything but the highest quality, evidence-based information on critical issues such as vaccination. Such techniques pose the danger that bad actors may take advantage of these coherent outputs as LLMs are not subject to truth or ethics (Bender et al., 2021).

Our examination of sources found levels of attribution similar to Kelly and colleagues (2023). Unlike some prior work (Liu et al., 2023; Malaviya et al., 2023; Shi et al., 2023), we did not assess the validity of the specific sources or the extent to which they support claims made. Nevertheless, our domain-level analysis offers valuable insights. Notable is the heavy dependence on news and media content across systems and topics. This highlights the complex relationship between media organizations and this new generation of search engines, which no longer function as directories, but consume media content to produce an information service. More surprising is the prominence of business websites, particularly in Perplexity responses. This evident commercial bias may be an outcome of search engine optimization and the practice of publishing blogs and articles on company websites to increase ranking and traffic. Equally, this may be an intentional step towards a form of native advertising (Zelch et al., 2024) and a potential revenue model. Either way, both systems seemed to interpret the renewable energy topic as commercial in intent, which was not our interpretation in designing the study. In addition to commercial bias, we found evidence of geographic bias in the heavy reliance on sources from the U.S. and other western countries. On a practical level, the low number of sources from Canada is surprising, given the queries were issued in Canada. However, the extent to which these systems employ geographic personalization is not known. Bing Chat did provide more results for the vaccination query from Canada, which suggests some use of geolocation data. More broadly, it will be important to determine the extent to which regional and linguistic bias places limits on these systems.

With respect to quality, while many of the cited sources cited are appropriate and authoritative, the value of these systems is seriously undermined by the seemingly random inclusion of poor-quality sources of information, including commercial blogs, clickbait, essay mills, and content farms. Yes, the Internet is full of such content, but given the nature of the topics in this study, such sources are wholly inappropriate. If they were retrieved as one item in a ranked list, they could be easily assessed and passed over, but with GenAI SEs, each source is integral to the response, so the quality of each resource must be held to a higher standard.

Finally, we found unmistakable evidence of design differences between these systems. Bing Chat responses are shorter, more conversational, easier to read, more positive in sentiment, and results are drawn from a narrower set of domains in comparison to Perplexity. Perplexity uses a more academic tone and approach, and the reading level is set higher than other available chatbots (Roegiest & Pinkosova, 2024). At the same time, it is more U.S.-centric and commercial in orientation, at least for these topics. The subtle differentiation of these tools points towards a path in which search engines will become even more adaptive, and hyper-personalized.

## CONCLUSION

This study makes several contributions to this rapidly developing area of research on GenAI SEs. Following best practices for algorithmic system audits, we found evidence of sentiment bias in relation to queries and topics and commercial and geographic bias in the cited sources. The different profiles adopted by Bing Chat and Perplexity highlight the extent to which GenAI SEs can be tailored and potentially influential in shaping public opinion in different directions. Responses clearly seek to establish cognitive authority by making claims and providing evidence, and yet the strength of these claims are undermined by ambiguous language and sources of questionable quality. We acknowledge the limitations of this study and the need for further work. We focused on 4 broad topics of public importance at one point in time (June 2023), and future work will need to examine a wider range of topics and tasks to claim evidence of general patterns. The orientation of the study was to consider the human impacts of these systems, in keeping with the conference theme of "people first," but our method did not involve human participants. In our future work, we will design user studies to test and validate these findings in relation to authority, credibility, and source quality.

## GENERATIVE AI USE

The outputs of Generative AI systems were the subject of this study. We did not use them in any other steps of analysis or writing. The authors assume all responsibility for the content of this submission.

## ACKNOWLEDGMENTS

We thank Shawn Ching-Chung Hsueh and Jason Li for their support in the data collection and analysis stages. This work is supported in part by funding from the Social Sciences and Humanities Research Council of Canada (SSHRC) CGS-D program, the Natural Sciences and Engineering Research Council of Canada (NSERC) CREATE program, and the Anne and George Piternick Student Research Award from the School of Information, University of British Columbia.

## REFERENCES

AI team. (2023, August 25). Performance parity with GPT-4. *Perplexity.* https://www.perplexity.ai/hub/blog/copilot-on-perplexity

Arazy, O., Kopak, R., Hadar, I. (2017). Heuristic principles and differential judgments in the assessment of information quality. *Journal of the Association for Information Systems, 18*(5), 403-432. https://doi.org/10.17705/1jais.00458

Azzopardi, L. (2021). Cognitive biases in search: A review and reflection of cognitive biases in information retrieval. *Proceedings of the 2021 Conference on Human Information Interaction and Retrieval*, 27–37. https://doi.org/10.1145/3406522.3446023

Baeza-Yates, R. (2018). Bias on the web. *Communications of the Association of Computing Machinery (ACM), 61*(6), 54–61. https://doi.org/10.1145/3209581

Bansal, S. & Aggarwal, C. (2022, March 15). textstat 0.7.3. PyPI. https://pypi.org/project/textstat/ .*GitHub.* https://github.com/pln-fing-udelar/fast-krippendorff

Bender, E. M., Gebru, T., McMillan-Major, A., & Shmitchell, S. (2021). On the dangers of stochastic parrots: Can language models be too big? 🦜. *Proceedings of the 2021 ACM Conference on Fairness, Accountability, and Transparency*, 610–623. https://doi.org/10.1145/3442188.3445922

Bing. (2023, February). *Bing conversational experiences and image creator terms.* https://www.bing.com/new/termsofuse/archives?features=bingnewpage

Boykoff, M.T. & Boykoff, J.M. (2004). Balance as bias: global warming and US prestige press, *Global Environmental Change, 14*(2), 125-136. https://doi.org/10.1016/j.gloenvcha.2003.10.001

Bubeck, S., Chandrasekaran, V., Eldan, R., Gehrke, J., Horvitz, E., Kamar, E., Lee, P., Lee, Y. T., Li, Y., Lundberg, S., Nori, H., Palangi, H., Riberio, M. T., & Zhang, Y. (2023). Sparks of artificial general intelligence: Early experiments with gpt-4. *arXiv preprint.* https://doi.org/10.48550/arXiv.2303.12712

Capra, R., & Arguello, J. (2023). How does AI chat change search behaviors? *arXiv preprint.* https://doi.org/10.48550/arXiv.2307.03826

Castro, S. (2023). Fast Krippendorff. *Pypi.* https://pypi.org/project/krippendorff/

Demartini, G., & Siersdorfer, S. (2010). Dear search engine: What's your opinion about...?: sentiment analysis for semantic enrichment of web search results. *Proceedings of the 3rd International Semantic Search Workshop on - SEMSEARCH '10*, 1–7. https://doi.org/10.1145/1863879.1863883

Edelman. (2023). e*delman trust barometer: Navigation a polarized world.* https://www.edelman.com/trust/2023/trust-barometer.

Friedman, B., & Nissenbaum, H. (1996). Bias in computer systems. *ACM Transactions on Information Systems,14*(3), 330 –347. https://doi.org/10.1145/230538.230561

Ghezzi, P., Bannister, P. G., Casino, G., Catalani, A., Goldman, M., Morley, J., Neunez, M., Prados-Bo, A., Smeesters, P. R., Taddeo, M., Vanzolini, T., & Floridi, L. (2020). Online information of vaccines: Information quality, not only privacy, is an ethical responsibility of search engines. *Frontiers in Medicine*, 7. https://doi.org/10.3389/fmed.2020.00400

Government of Canada (2018, October 19). *The next generation of emerging global challenges.* https://horizons.gc.ca/en/2018/10/19/ the-next-generation-of-emerging-global-challenges/.

Haider, J. & Sundin, O. (2019). *Invisible search and online search engines: The ubiquity of search in everyday life* (1st ed.). Routledge.

Harris, C. R., Millman, K. J., Van Der Walt, S. J., Gommers, R., Virtanen, P., Cournapeau, D., Wieser, E., Taylor, J., Berg, S., Smith, N. J., Kern, R., Picus, M., Hoyer, S., Van Kerkwijk, M. H., Brett, M., Haldane, A., Del Río, J. F., Wiebe, M., Peterson, P., . . . Oliphant, T. E. (2020). Array programming with NumPy. *Nature, 585*(7825), 357–362. https://doi.org/10.1038/s41586-020-2649-2

Hilligoss, B., & Rieh, S. Y. (2008). Developing a unifying framework of credibility assessment: Construct, heuristics, and interaction in context. *Information Processing & Management, 44*(4), 1467–1484. https://doi.org/10.1016/j.ipm.2007.10.001

Hovland, C., Janis, I., & Kelley, H. (1953). *Communication and persuasion.* New Haven, GT: Yale University Press.

Huang, P. S., Zhang, H., Jiang, R., Stanforth, R., Welbl, J., Rae, J., ... & Kohli, P. (2020, November). Reducing sentiment bias in language models via counterfactual evaluation. *Findings of the Association for Computational Linguistics: EMNLP 2020* (pp. 65-83). https://doi.org/10.18653/v1/2020.findings-emnlp.7

Hussein, E., Juneja, P., & Mitra, T. (2020). Measuring misinformation in video search platforms: An audit study on YouTube. *Proceedings of the ACM on Human-Computer Interaction, 4(CSCW1)*, 1–27. https://doi.org/10.1145/3392854

Huvila, I. (2013). *In Web search we trust? Articulation of the cognitive authorities of Web searching.* https://informationr.net/ir/18-1/paper567.html

Ji, Z., Lee, N., Frieske, R., Yu, T., Su, D., Xu, Y., Ishii, E., Bang, J., Madotto, A. & Fung, P. (2023). Survey of hallucination in natural language generation. *ACM Computing Surveys, 55*(12), 1-38. https://doi.org/10.1145/3571730

Kay, M., Matuszek, C., & Munson, S. A. (2015, April). Unequal representation and gender stereotypes in image search results for occupations. *Proceedings of the 33rd annual ACM conference on human factors in computing systems* (pp. 3819-3828). https://doi.org/10.1145/2702123.2702520

Kazai, G., Thomas, P., & Craswell, N. (2019, July). The emotion profile of web search. *Proceedings of the 42nd international ACM SIGIR conference on research and development in information retrieval* (pp. 1097-1100). https://doi.org/10.1145/3331184.3331314

Kelly, D., Chen, Y., Cornwell, S. E., Delellis, N. S., Mayhew, A., Onaolapo, S., & Rubin, V. L. (2023). Bing Chat: The future of search engines? *Proceedings of the Association for Information Science and Technology, 60*(1), 1007–1009. https://doi.org/10.1002/pra2.927

Kincaid, J. P., Fishburne Jr, R. P., Rogers, R. L., & Chissom, B. S. (1975). Derivation of new readability formulas (automated readability index, fog count and flesch reading ease formula) for navy enlisted personnel. *Institute for Simulation and Training, University of Central Florida*. https://stars.library.ucf.edu/istlibrary/56/?utm_sourc

Krippendorff, K. (2019). Reliability. *Content analysis: An introduction to its methodology* (Fourth ed., pp. 277-360). SAGE Publications. https://doi.org/10.4135/9781071878781

Le, B., Spina, D., Scholer, F., & Chia, H. (2022). A crowdsourcing methodology to measure algorithmic bias in black-box systems: A case study with COVID-related searches. In L. Boratto, S. Faralli, M. Marras, & G. Stilo (Eds.), *Advances in bias and fairness in information retrieval* (Vol. 1610, pp. 43–55). Springer International Publishing. https://doi.org/10.1007/978-3-031-09316-6_5

Li, A., & Sinnamon, L. (2023, August 28- September 2). Examining query sentiment bias effects on search results in large language models. *The Symposium on Future Directions in Information Access (FDIA) co-located with the 2023 European Summer School on Information Retrieval (ESSIR)*. https://2023.essir.eu/FDIA/papers/FDIA_2023_paper_2.pdf

Li, D., Sun, Z., Hu, X., Liu, Z., Chen, Z., Hu, B., Wu, A., & Zhang, M. (2023). A survey of large language Models attribution. *arXiv preprint*. https://doi.org/10.48550/arXiv.2311.03731

Liu, N., Zhang, T., & Liang, P. (2023). Evaluating verifiability in generative search engines. *Findings of the Association for Computational Linguistics: EMNLP 2023*, 7001–7025. https://doi.org/10.18653/v1/2023.findings-emnlp.467

Mager, A., Norocel, O. C., & Rogers, R. (2023). Advancing search engine studies: The evolution of Google critique and intervention. *Big Data & Society, 10*(2). https://doi.org/10.1177/20539517231191528

Malaviya, C., Lee, S., Chen, S., Sieber, E., Yatskar, M., & Roth, D. (2023). ExpertQA: Expert-curated questions and attributed answers. *arXiv preprint*. https://doi.org/10.48550/ARXIV.2309.07852

McKinney, W. (2010). Data structures for statistical computing in Python. *Proceedings of the Python in Science Conferences*. https://doi.org/10.25080/majora-92bf1922-00a

Media Bias/Fact Check. (2024, April 8). *Grunge – bias and credibility*. https://mediabiasfactcheck.com/grunge/

Mehdi, Y. (2023, March 14). Confirmed: the new Bing runs on OpenAI's GPT-4. *Microsoft Bing Blogs*. https://blogs.bing.com/search/march_2023/Confirmed-the-new-Bing-runs-on-OpenAI%E2%80%99s-GPT-4

Metaxa, D., Gan, M. A., Goh, S., Hancock, J., & Landay, J. A. (2021a). An image of society: Gender and racial representation and impact in image search results for occupations. *Proceedings of the ACM on Human-Computer Interaction, 5*(CSCW1), 1–23. https://doi.org/10.1145/3449100

Metaxa, D., Park, J. S., Robertson, R. E., Karahalios, K., Wilson, C., Hancock, J., & Sandvig, C. (2021b). Auditing algorithms: Understanding algorithmic systems from the outside in. *Foundations and Trends® in Human–Computer Interaction, 14*(4), 272-344. http://dx.doi.org/10.1561/1100000083

Noble, S. U. (2018). *Algorithms of oppression: How search engines reinforce racism.* New York University Press.

NP Digital. (2023). Discover what people are asking about... *Answer the Public.* https://answerthepublic.com/

O'Connor, C., & Joffe, H. (2020). Intercoder reliability in qualitative research: debates and practical guidelines. *International Journal of Qualitative Methods*, *19*, 1609406919899220. https://doi.org/10.1177/160940691989922

OpenAI. (2023a, March 14). *Terms of use.* https://openai.com/policies/mar-2023-terms

OpenAI. (2023b). *What is the ChatGPT Plus model selector?* https://help.openai.com/en/articles/7864572-what-is-the-chatgpt-plus-model-selector

Perplexity. (2023). *Perplexity's terms of service.* https://www.perplexity.ai/hub/legal/terms-of-service

Pichai, S. (2023, February 6). An important next step on our ai journey. *Google.* https://blog.google/intl/en-africa/products/explore-get-answers/an-important-next-step-on-our-ai-journey/

Python Software Foundation. (2023). *re — regular expression operations, release 3.10.12.* https://docs.python.org/3.10/library/re.html.

Rieh, S. Y. (2002). Judgment of information quality and cognitive authority in the Web. *Journal of the American society for information science and technology, 53*(2), 145-161.

Roegiest, A., & Pinkosova, Z. (2024, March). Generative information systems are great if you can read. *Proceedings of the 2024 ACM SIGIR Conference on Human Information Interaction and Retrieval* (pp. 165-177). https://doi.org/10.1145/3627508.3638345

Rubin, V. L. (2019). Disinformation and misinformation triangle: A conceptual model for "fake news" epidemic, causal factors and interventions. *Journal of Documentation, 75*(5), 1013-1034. https://doi.org/10.1108/JD-12-2018-0209

Sandvig, C., Hamilton, K., Karahalios, K., & Langbort, C. (2014). Auditing algorithms: Research methods for detecting discrimination on internet platforms. *Data and discrimination: converting critical concerns into productive inquiry, 22,* 4349-4357.

Shah, C., & Bender, E. M. (2022). Situating search. *ACM SIGIR Conference on Human Information Interaction and Retrieval, 221–232*. https://doi.org/10.1145/3498366.3505816

Shen, H., DeVos, A., Eslami, M., & Holstein, K. (2021). Everyday algorithm auditing: Understanding the power of everyday users in surfacing harmful algorithmic behaviors. *Proceedings of the ACM on Human-Computer Interaction, 5(CSCW2)*, 1–29. https://doi.org/10.1145/3479577

Shi, X., Liu, J., Liu, Y., Cheng, Q., & Lu, W. (2023). Know where to go: Make LLM a relevant, responsible, and trustworthy searcher. *arXiv preprint.* https://doi.org/10.48550/arXiv.2310.12443

Smith, M. S. (2024, February 24). Perplexity.ai Revamps Google SEO Model For LLM Era > AI search leader mixes Meta-built smarts with scrappy startup fervor. *IEEE Spectrum.* https://spectrum.ieee.org/perplexity-ai

Sundin, O. (2011). Janitors of knowledge: Constructing knowledge in the everyday life of Wikipedia editors. *Journal of Documentation, 67*(5), 840–862. https://doi.org/10.1108/00220411111164709

Taraborelli, D. (2008). How the Web is changing the way we trust. In A. Briggle, K. Waelbers, & P. A. E. Brey (Eds.). *Proceeding of the 2008 Conference on Current Issues in Computing and Philosophy* (pp. 194-204) Amsterdam: IOS Press.

Thomas, D. R. (2006). A general inductive approach for analyzing qualitative evaluation data. *American journal of evaluation*, *27*(2), 237-246. https://doi.org/10.1177/1098214005283748

Tian, J., Chen, S., Zhang, X., Wang, X., & Feng, Z. (2023, June). Reducing sentiment bias in pre-trained sentiment classification via adaptive gumbel attack. *Proceedings of the AAAI Conference on Artificial Intelligence, 37*(11), 13646-13654. https://doi.org/10.1609/aaai.v37i11.26599

Tripodi, F. (2018). Searching for alternative facts. *Data & Society.* https://datasociety.net/wp-content/uploads/2018/05/Data_Society_Searching-for-Alternative-Facts.pdf

Urman, A., & Makhortykh, M. (2023). The silence of the LLMs: Cross-lingual analysis of political bias and false information prevalence in ChatGPT, Google Bard, and Bing Chat. *OSF Preprints.* https://doi.org/10.31219/osf.io/q9v8f

Wazzan, A., MacNeil, S., & Souvenir, R. (2024). Comparing traditional and LLM-based search for image geolocation. *Proceedings of the 2024 ACM SIGIR Conference on Human Information Interaction and Retrieval* (pp.291-302). https://doi.org/10.1145/3627508.3638305

White, R. W. (2013, July 28-August 1). Beliefs and biases in web search. *Proceedings of the 36th international ACM SIGIR conference on Research and development in information retrieval* (pp. 3-12). https://doi.org/10.1145/2484028.2484053

White, R. W. (2024). Advancing the search frontier with AI agents. *arXiv preprint.* https://doi.org/10.48550/arXiv.2311.01235

Wilson, P. (1983). *Second-hand knowledge: An inquiry into cognitive authority.* Westport, CT: Greenwood Press.

Yuan, L. X. (2023). Distilbert-base-multilingual-cased-sentiments-student. *Hugging Face.* https://huggingface.co/lxyuan/distilbert-base-multilingual-cased-sentiments-student.

Zelch, I., Hagen, M., & Potthast, M. (2024). A user study on the acceptance of native advertising generative IR. *Proceedings of the 2024 ACM SIGIR Conference on Human Information Interaction and Retrieval* (pp. 142-152). https://doi.org/10.1145/3627508.3638316